# Decentralization as Organizing Principle of Emergent Urban Structures[1]

**Rainer E. Zimmermann**
IAG Philosophische Grundlagenprobleme,
FB 1, UGH, Nora-Platiel-Str.1, D – 34127 Kassel /
Clare Hall, UK – Cambridge CB3 9AL[2] /
Lehrgebiet Philosphie, FB 13 AW, FH,
Lothstr.34, D – 80335 München[3]
e-mail: pd00108@mail.lrz-muenchen.de

## Abstract

With a view to the ongoing Bologna project (www.arXiv.org, nlin.AO/0109025) general organizing principles of emergent structures in social systems are being discussed with a view to the meaning of decentralization. It is proposed to introduce decentralization as a principle for organizing emergent structures in a generic way utilizing aspects of the insight gained by the Santa Fe school dealing with self-organized criticality. The techniques utilized come from graph theory, category theory, and in particular quantum gravity, which bear a strong potential for a multitude of applications in research fields with a significant interdisciplinary scope. This is especially important for applications in the organization of social systems which usually call for an interaction of logic and hermeneutic.

## 1  Introduction: The Mediation of Micro- and Macrolevels

When dealing with the emergence of structures, the *phenomenon* of emergence itself, and the *concept* of emergence we derive from it, point mainly to a systematic deficiency of inference. In other words, on the one hand, emergence refers to models rather than to processes: Very much like the concept of complexity, the concept of emergence is attributed to the models, not to processes proper. The reason for this is that we never actually deal with the „real" world as it is (in ontological terms), but only with the „modal" world as we model it (in epistemological terms) according to what we perceive. On the other hand, emergence is closely related to innovation. We call a phenomenon *innovative*, if its appearance cannot be inferred within a utilized language describing some process level, although the language is completely specified. To be more precise: We cannot formulate a model in a macrolanguage, given the microlanguage. ([5], p.

---

[1] Talk delivered at the AKSOE meeting, section 14, on the occasion of the 2002 spring conference of the DPG, Regensburg.
[2] Permanent addresses.
[3] Present address.

47, cf. p.72) The point is here that we visualize processes as acting on two levels relative to each other, one the *microlevel* on which a large number of relevant processes take place being performed by individual agents associated with specific actions, the other the *macrolevel* on which a collective phenomenon can be observed which can be understood as a kind of superposition of all the individual actions being performed on the microlevel. In other words: Observable processes show up on the macrolevel of he world, this level being defined primarily in terms of the perceptive properties of observers. But they „happen" on the microlevel. This is basically true for all possible systems. But in the case of *social* systems, the agents on the microlevel and the observers on the macrolevel can be identical. This introduces an explicit notion of *self-reference* into the interpretation of observed phenomena. The result is that detailed behaviour of individuals who comprise a given social aggregate cannot be retraced in the observed collective behaviour of that same aggregate. This aspect has been discussed in detail by Thomas Schelling [18], but much earlier already by Henri Lefebvre [13] and Jean-Paul Sartre [17] who term it „counter-finality". With a view to social systems of this kind, self-reference shows up in the fact that on the microlevel the „purposive" behaviour of persons is a *relational* concept in itself, because it depends on constraints determined by the environment which reduce its degrees of freedom eventually rendering it completely contingent. On the other hand, individuals depend on their essentially *local* knowledge produced by *local* interactions (on their microlevel) so that the *global* outcome is generically a result of superposition, not strictly independent of a single action taken, but in any case *different* from it. Institutional constraints (e.g. of legal sort), and feedback loops (of information) complicate the mediation of micro- and macrolevels and endow the process with an implicitly *recursive* characteristic. ([18], pp. 20sq., 33, 50) There is another point to this: We realize that social processes depend decisively on the concrete information flow. In this sense, a process (or phenomenon) can be said to be emergent, if at some time the architecture of information processing has changed such that a more powerful level of intrinsic computation has appeared which has not been present before. This criterion taken from Crutchfield [2] can be shown to be essentially equivalent to the mediation criterion between microlanguage and macrolanguage given above. Visualized this way, innovation occurs at the threshold of information processing when the agent's modelling capacity approaches the complexity of the same agent's *internal* model. ([2], p.25) This is so because (as we have seen above) the agent is always co-acting with others, and thus each model contains at least one self-model. At the same time, this has an important consequence for the relationship between order and disorder: because *stability* (order) is necessary for the consistent information storage while *instability* (disorder) is necessary for the production of information and its communication. Hence, it is the dialectic of stability and instability which is characterizing the modelling of evolutionary hierarchies of structure. And it is this trade-off which is relevant for computation theory. So if agents model their environment, their results depend on their capacity of identifying regularities (which explicitly takes place in terms of differentiating differences). This in turn depends on their computational resources and the language (model classes) utilized. Innovative emergence then, is the representation of attaining a new model class which improves the modelling process in the first place.

What we find therefore, is that the production of knowledge in terms of modelling the environment for the benefit of an adequate inference about observable processes can be visualized as an algorithmic form of compuational process. This does not only mean that social processes are being modelled in terms of computational processes, but that in fact they *are* such processes. In other words: Information is physical. „Our world is the information process that is running in the computer, but this computer is not in our world." (Edward Fredkin) ([24], p.58) In this sense, information shows up as the foundation of the reality we perceive. This computational process is *algorithmic* in the sense that for social systems, „purposive" behaviour can be reflected by strategic choices of actions undertaken. For humans particularly, but not exclusively, the inventory of available algorithms is explicitly determined by the process of socialization. Hence, when talking about computation in this very fundamental sense, we are referred to a methodological framework which is concerned with problems of graph colouring, connectivity of networks, and so forth. Indeed, combinatorial optimization can be applied to such problems, despite the problem of irrational action, because information within this context can be roughly compared to „abstract pheromone" as it is introduced in algorithmic studies of artificial ants colonies. ([4], [25])

Now, the topic of *decentralization* is important for all of this, because it is reflecting a basic organizational structure which governs both the phenomenological level of observations and the methodological level of modelling, respectively. In terms of the latter, this can be seen as a result of the intrinsic multiperspectivity of the models utilized. Hence, decentralization turns out to be a unified *onto-epistemic* concept rather than a mere pragmatic aspect. [16] In terms of the former, decentralization shows up as a straightforward consequence of deriving global coordination from local interactions. As Crutchfield and Mitchell have discussed in detail, centralized organization has three major drawbacks as compared to decentralized organzation: namely with respect to the *speed of information processing* (centres can be bottlenecks), *robustness* (central failure destroys system structure), and *equitable resource allocation* (centres draw resources). [3] Crutchfield and Mitchell claim (and we agree) that evolution proceeds via a series of epochs which are connected to each other by distinct computational innovations. Hence, they can visualize the problem illustrated here as the evolutionary discovery of methods for emergent global computation in a spatially distributed system consisting of locally interacting processors. This is a useful viewpoint to be taken when talking about the simulation of such processes. And there are interesting results dealing with what they call „propagation of domain walls" (or embedded particles). Note however that for the pro-blem we have in mind here when referring to the emergence and evolution of urban structures, the *concepts of space and time* themselves have to be questioned. Because social structures, such as urban structures in particular, say, are not actually evolving within a given space: Instead, they *produce* their own space. [13] This aspect has been recently discussed in more detail (unfortunately, without referring to the pioneering work of Lefebvre) by Hillier and Hanson. ([8], [9]) In order to model this primarily architectural space they develop what they call a „morphic language" different from mathematical modelling in the strict sense, and from natural languages utilized in hermeneutics: In fact, while the latter strongly individuates primary morphic units within the framework of a comparatively permissive formal (syntactic) structure, the former utilizes very small lexicons and very large syntaxes which are virtually useless for representing the world as it appears. Hence, visualized this way, mathematical languages do not represent or mean anything except their own structure. The morphic language however utilizes a small lexicon and gives primacy of syntax over semantic representation. Hence, it is built up from a minimal initial system meaning its own structure (similar to mathematics), but it can be realized within experience and is subject to a rule-governed sort of creativity (similar to natural languages). ([9], pp. 48-50) For describing the intelligibility of space, Hillier and Hanson explicitly utilize aspects of combinatorial topology when modelling spatial distributions of urban structures, often referring to (information) transport properties of graphs, connectivity, and percolation. This is actually the point from where we start our own approach utilizing a more generalized conception from the outset.

The idea is to eventually draw on the resources of category theory (or topos theory), because this is the field which comes closest to verbal conceptualization in the sense of Hillier's and Hanson's morphic language, without giving up the precise syntactic of mathematics. The point is mainly that categories provide both: the mathematical modelling of physically observable structures, and the logic conceptualization of this modelling, at the same time. This is the reason for starting here with relevant aspects of fundamental physics (in terms of recent results on quantum gravity). They may serve as a general framework for ultimately translating modelling procedures from physics to social science, avoiding the complete retreat into the domain of qualitative metaphors, but furnishing instead a semi-quantitative (morphic in this sense) approach which is resilient enough to remain relevant and sufficiently stable on the edge between logic and hermeneutic. However, it is not possible here to assemble all aspects of the theory utilized. This has been done in more detail elsewhere. (Cf. [27] through [35] for more explicit expositions. In particular, [32] gives a recent, not too technical survey of the general idea as derived from physics including detailed references. See [36] for an introduction to the Bologna project where these aspects shall be put to empirical test.) But what we will do here is to shortly summarize the basic ideas, discussing emergent computation as implicit in the concept of spin networks introduced by Roger Penrose (section 2), and referring to the underlying fundamental aspects of this approach as they are relevant for our topic here (section 3). We will soon notice that presently ongoing work in this field turns out to be not far removed from the line of approach taken here, although the methods applied may be diffferent in detail. But with a view to more recent expositions on complexity theory or population ecology such as [10] and [23], and on spatial economy such as [11] and [7], we can recover a lot of what will be said here under a more fundamental perspective. This is also true for a number of practical issues put forward in the political field. (See [1], [6], [12], [14], [15]. Also [25] is very illustrative for economic issues.) Similar aspects can also be retrieved from [19] through [22]. But first we start with the already mentioned point of visualizing space and time, the basic categories in which we perceive and reflect, themselves, as derived (hence produced) quantities which constitute our world as we perceive it.

## 2  Emergent Computation

The original idea of introducing a purely combinatorial (abstract) structure in order to eventually „derive" space and time from it such that relativity and quantum theory show up as two different perspectives of the same underlying whole, goes back to Roger Penrose who invented *spin networks*. Basically, spin networks in this sense are trivalent graphs with a combinatorial loading called *spin numbers*. The idea is to represent abstract interactions between vertices: Be a, b, c natural numbers such that for any i, j, k the conditions $i + j = a$, $i + k = b$, $j + k = c$ can be solved with non-negative integers. Then the edges of the graph can be visualized as expressing interactions of particles with spin a and b which produce a particle with spin c. Hence, for an unrestricted number of vertices and edges, we get a spin network consisting of vertices interacting by permanently exchanging spin numbers n such that $n = 2s$ („measured" in terms of ℏ). We could visualize therefore, the network as one which is generically fluctuating. Note however that *no conception* of space and time is entering the definition yet, so that we deal with a purely combinatorial process producing numbers as an output according to a given input. In this sense, we can think of this network fluctuation as of a computational process. In other words: There is a *permanent processing of information* which is underlying the structure of space and time on a fundamental level. If we take „large portions" of this network then, and choose two single strands of it, we can define the following relational structure: Take large portions K of the network, single out two smaller strands, N and M say, and compare their relative positions in terms of an isolated single strand (acting as a single edge) of spin number 1, then we can, according to the possibilities of re-combination, determine an appropriate *angle between units* which by superposition produces an angle in terms of Euklidean geometry. The total loading of a unit is given by what Penrose calls *value* of the network, essentially expressing a scalar product of spin network states. Be Γ such a state, and Γ* its dual. Think of the portions of the network as being knotted by closing the strands to them-
selves. Then, in Dirac notation, we have for the value V of a spin network

$$<\Gamma \,|\, \Gamma> = V(\Gamma \# \Gamma^*)$$

with

$$V[...] = \prod (1/j!) \sum \varepsilon (-2)^N.$$

Here, the product is being taken over all edges, and the sum over all vertices.
Then, j is the respective edge label, N the number of closed loops, and ε the
permutation rule for the signs being referred to as *intertwining operation*. The
*spin geometry theorem* tells us then that the probability ratio of finding the net-
work in consecutive states (by e.g. repeating the procedure of comparing isola-
ted strands with respect to the „lump" network twice) is given by (1/2) cos θ,
with θ being a (Euklidean) angle. Hence, by a statistical choice of strands from
the (lump) network, we can define an angle and thereby actually *produce* it out
of a purely combinatorial structure due to a combinatorial process of selection.
In this sense, the fundamentally underlying process of computation described
here in preliminary terms as a permanent processing of information, is also a
process of *emergent computation* such that from the underlying process classi-
city can eventually emerge and show up in terms of macroscopic physical
structures. The network structure can be made more explicit when projecting it
onto a spherical surface. In fact, as it turns out, spin networks can be better vi-
sualized as a fundamental combinatorial level of *space* rather than of space-time
as originally intended by Penrose. In this sense, Rovelli and Smolin have re-
interpreted his ansatz somehow: They start with loops from the outset and show
that since spin network states <S | span the loop state space, it follows that any
ket state |ψ> is uniquely determined by the values of the S-functionals on it,
namely of the form ψ(S) := <S | ψ>. To be more precise, Rovelli and Smolin ta-
ke embedded spin networks rather than the usual spin networks, i.e. they take
the latter plus an immersion of its graph into a 3-manifold. Considering then, the
equivalence classes of embedded oriented spin networks under diffeomor-
phisms, it can be shown that they are to be identified by the knotting properties
of the embedded graph forming the network and by its colouring (which is the
labelling of its links with positive integers referring to spin numbers). When ge-
neralizing this concept even further, a network design may be introduced as a
conceptual approach towards *pre-geometry* based on the elementary concept of
*distinctions*, as Louis Kauffman has shown. In particular, space-time can be vi-
sualized as being produced directly from the operator algebra of a distinction. If
thinking of distinctions in terms of 1-0 (or *yes-no*) decisions, we have a direct
link here to information theory, which has been discussed recently again with a
view to holography. The network which is the dual of the spin network is the appropriate triangulation of the spherical surface. This triangulation forms the dual 1-skeleton of the spin network. The length numbers attributed to triangle edges correspond to the spin numbers in the original network. Hence, they change accordingly. As the triangulation is the minimal covering of this surface, the length of a triangle edge gives a quantized portion of (3-) space. In other words, the quantum of a surface (area) and the quantum of a space (volume) can be derived directly from the network pattern. To be more precise, area A and volume V will turn out to be proportional to $l_p^2$ and $l_p^3$, respectively. So we can think of the „fundamental level" of physics (or rather of the boundary layer of this level) in terms of the fluctuating network structure corresponding at the same time to a triangulation which is the network's dual. If one visualizes this structure as one which is fluctuating all the time, then one gets an animated impression of the fundamental level of physics which is easily comparable with a kind of permanent computational process underlying all what there actually is. This is roughly comparable to permanently ongoing communication on the microlevel of society.

Following a convention introduced by Baez, the permanent re-arrangement of spin numbers (and triangle edge lengths) is called *spin foam*. Note that visualized as a sequence, a spin foam can be thought of as a kind of time evolution of spin networks. However, the concept of time (and causality) is ill-defined on that fundamental level. A straightforward „quantization" of time (proportional to $t_p$) is not satisfactory so far, because it depends on the outcome of the question whether there is any time ordering on that level at all and whether therefore, time is an emergent concept (as is ultimately space). But if visualized as a sequence of spin network states, spin foams (thought of as constituting a kind of proto-space-time) turn out to be two-dimensional analogues of Feynman graphs: They are in fact 2-complexes with faces, edges, and vertices such that the amplitudes of faces and edges correspond to propagators and vertex amplitudes to interactions. A very useful model for spin foams is given by Barrett and Crane (developed 1998 until 2000) as a good candidate for illustrating the basic idea. The amplitude of a spin foam is given by the following product:

$$Z(F) = \prod_{f \in \Delta(2)} A(f) \prod_{e \in \Delta(3)} A(e) \prod_{v \in \Delta(4)} A(v),$$

where f, e, v are faces, edges, and vertices, respectively, and the $\Delta(n)$ are the n-simplices in the triangulation. The partition function is then

$$Z(M) = \sum_F Z(F).$$

Now what is the „macroscopic" equivalent of the motion described here in terms of the „microscopic" level? Recall that according to the standard terminology, a *loop* in some space $\Sigma$, say, is a continuous map $\gamma$ from the unit interval into $\Sigma$ such that $\gamma(0) = \gamma(1)$. The set of all such maps will be denoted by $\Omega\Sigma$, the loop space of $\Sigma$. Given a loop element $\gamma$, and a space $A'$ of connections, we can define a complex function on $A' \times \Omega\Sigma$, the socalled *Wilson loop* such that $T_A(\gamma) :=$ (1/N) Tr$_R$ P exp $\int_\gamma A$. Here, the path-ordered exponential of the connection $A \in A'$, along the loop $\gamma$, is also known as the holonomy of A along $\gamma$. The holonomy measures the change undergone by an internal vector when parallel transported along $\gamma$. The trace is taken in the representation R of G (which is actually the Lie group of Yang-Mills theory), N being the dimensionality of this representation. *The quantity measures therefore the curvature* (or field strength) *in a gauge-invariant way*. Over a given loop $\gamma$, the expectation value $< T(\gamma)>$ turns out to be equal to a *knot invariant* (the „Kauffman bracket") such that when applied to spin networks, the latter shows up as a deformation of Penrose's value V($\Gamma$). Hence, spin networks are deformed into *quantum* spin networks (which are essentially given by a family of deformations of the original networks of Penrose). There is also a simplicial aspect to this: Loop quantum gravity provides for a quantization of geometric entities such as area and volume. The main sequence of the spectrum of area e.g., shows up as A = $8\pi\gamma\hbar$G $\sum_i(j_i(j_i + 1))^{1/2}$, for c = 1, where the j's are half-integers labelling the eigenvalues. (Compare this with the remark on space quantization above.) This quantization shows that the states of the spin network basis are eigenvalues of some area and volume operators. We can say that a spin network carries quanta of area along its edges, and quanta of volume at its vertices. A quantum space-time can be decomposed therefore, in a basis of states visualized as made up by quanta of volume which in turn are separated by quanta of area (at the intersections and on the links, respectively). Hence, we can visualize a spin network as sitting on the dual of a cellular de-composition of physical space. As far as the dynamics of spin networks is concerned, there is still another, more recent approach, which appears to be promising as to the further development of topological aspects of quantum gravity (referred to as TQFT).

We notice from what we said above that a spin foam is a two-dimensional complex built from vertices, edges, and polygonal faces, with the faces labelled by group representations, and the edges labelled by intertwining operators. If we take a generic slice of a spin foam, we get a spin network. Hence, a spin foam is essentially taking one spin network into another, of the form F: $\Psi \rightarrow \Psi'$. Just as spin networks are designed to merge the concepts of quantum state and geometry of space, spin foams shall serve the merging of concepts of quantum history and geometry of space-time. Very much like Feynman diagrams do, also spin foams provide for evaluating information about the history of a transition of which the amplitude is being determined. Hence, if $\Psi$ and $\Psi'$ are spin networks with underlying graphs $\gamma$ and $\gamma'$, then any spin foam F: $\Psi \rightarrow \Psi'$ determines an operator from $L^2(A_\gamma /G_\gamma)$ to $L^2(A_{\gamma'} /G_{\gamma'})$ denoted by O such that $<\Phi', O \Phi> = <\Phi', \Psi'><\Psi, \Phi>$ for any states $\Phi$, $\Phi'$. The evolution operator Z(M) is a linear combination of these operators weighted by the amplitudes Z(O). This leads to a discrete version of a path integral. Hence, re-arrangement of spin numbers on the „combinatorial level" is equivalent to an evolution of states in terms of Hilbert spaces in the „quantum picture" and effectively changes the topology of space on the „cobordism level". This is the level we ultimately perceive when observing phenomena. In fact, change of *form* (Gestalt) can be visualized as a change of its underlying topology. Evolution (in macroscopic terms) can be understood therefore, as a kind of *manifold morphogenesis* in (emergent) time: Visualize the n(= 4, say)-dimensional manifold M (with $\partial M = S \cup S'$ - disjointly) as M: S $\rightarrow$ S', that is as a process (or as time) passing from an (n-1)-dimensional space S to another (n-1)-dimensional space S'. (Here n-1 = 3.) This is the mentioned *cobordism*. Note that composition of cobordisms holds and is associative, but *not commutative*. The identity cobordism can be interpreted as describing a passage of time when topology stays constant. Visualized this way, TQFT might suggest that general relativity and quantum theory are not so different after all. In fact, *the concepts of space and state turn out to be two aspects of a unified whole, likewise space-time and process*. So what we have in the end, is a rough (and simplified) outline of the foundations of emergence, in the sense that we can localize the fine structure of emergence (the re-arrangements of spin numbers in purely combinatorial terms being visualized as a fundamental fluctuation) and its results on the „macroscopic" scale (as a change of topology being visualized by physical observers). This is actually what we would expect of a proper *theory of emergence*. And this is also what we aim at in this present approach: The topological domain of the macroscopic scale will be discussed in terms of a suitable morphic language, if developed for discussing the emergence of social, e.g. urban structures. The underlying microscopic process structure which actually produces the macroscopic scale refers then to the actions taken by individual agents self-organizing into groups and institutions. In other words: What Hillier and Hanson call the intrinsic logic of an emergent spatial structure („space" here in the sense of „social space") is nothing but the operation of the selected algorithmic process representing the communicative ineractions on the microlevel of society. This operative procedure can be easily visualized as a process of emergent computation as motivated above.

## 3   Fundamental Information Processing

The important point is to notice here that from the beginning on, the concept of *time* is integrated into the geometrical representation from the outset. But it does not show up as a function, because the map which carries one space into the other is itself a manifold space. Hence, time is related to the frequency of topology changes encountered (by literally counting them). This is a notion of time which is very much on the line of the time concept introduced by Prigogine in the seventies. Essentially, time can be visualized as a measure for the structural change of the (spatial) world. These results can also be formulated in the lan-guage of category theory: As TQFT maps each manifold S representing space to a Hilbert space Z(S) and each cobordism M: S → S' representing space-time to an operator Z(M): Z(S) → Z(S') such that composition and identities are preserved, this means that TQFT is a functor Z: nCob → Hilb. Note that the non-commutativity of operators in quantum theory corresponds to non-commutativity of composing cobordisms, and adjoint operation in quantum theory turning an operator A: H → H' into A*: H' → H corresponds to the operation of reversing the roles of past and future in a cobordism M: S → S' obtaining M*: S' → S. So what we do realize after all is that spin networks, in particular as being visualized in terms of their quantum deformations, turn out in these models as the fundamental fabric of the world, in the sense that they are underlying and eventually actually producing the world of classical physics. Note in passing that this has an important philosophical consequence, because in being a functor and a theory *at the same time*, TQFT is constituted in an intrinsically *onto-epistemic* way: In other words, we model a physical structure in mathematical terms (with ontologically relevant results), and at the same time we model *the modelling itself* with respect to the specific logic which is underlying the category we are handling (with epistemologically relevant results. Hence, we do both: theoretical research and its conceptualizing.

The question is now as to the mediating concept which translates processes on the microlevel to processes on the macrolevel. This can be achieved by utilizing knot theory: We start for simplicity with crossings in non-oriented entanglements of strings. It is possible then to produce two diagrams out of a crossing by splitting it in two different ways. Labelling the various regions (produced by the first or the second version of splitting) by A and B, respectively, we find a whole family tree of possible splittings which end in a collection of Jordan curves called *states* of the diagram. Note that this tree structure exhibits clearly the relationship of form (Gestalt) and (underlying) dynamics: The „moves" necessary to resolve a knot and reduce the various crossings to states represent the *implicit* motion which is hidden underneath the morphological form of a knot. In other words, by performing prescribed (formal) moves one is actually *reconstructing the motion* which was necessary to form the knot in the first place. There is in fact a large class of possible (formalized) moves of this type, ranging from Reidemeister moves to Pachner moves (which turn out to be very significant with a view to the unfolding of spin networks, but on which we cannot comment here in this present paper). This relationship between actual „form" (morphology) and implicit motion (having eventually produced this form) resembles closely the relationship between implicit and explicit *order* as introduced a long time ago by David Bohm.  It is also comparatively easy to demonstrate the algebraic relevance of crossings – illustrating the aspect of implicit motion under another perspective: Let a crossing represent the interaction of two curves, of *a* and *b*, say. We can interpret the result of this interaction as an abstract product *ab* which turns out to be non-associative. Hence, a close relations-hip can be easily demonstrated between the graphical representation of (knot) crossings and a non-associative algebra (of knots). In fact, as Kauffman can show in more detail, this approach leads to logical structures, if the crossings are oriented. We associate now a polynomial with a knot diagram in the following way: If $|\sigma|$ is the number of loops in state $\sigma$ (after disentangling all crossings of a knot) and $<K|\sigma>$ the product of all markings (labellings) in $\sigma$, then the expression

$$<K> = <K>(A,B,d) = \sum_\sigma <K|\sigma> d^{|\sigma|}$$

gives a weighted sum over states and is called the *bracket polynomial* of the diagram K. We can also introduce a normalization condition in order to ensure ivariance of the polynomial under neighborhood isotopy. If so, then

$$L_K := (-A^3)^{-\omega(K)} <K>,$$

where $\omega(K) = \sum_{p \in C(K)} \varepsilon(p)$ is the winding number (permuting signs). In order to compare this with more well-known aspects of combinatorial topology, we can think of the appropriate colouring of knots which leads to problems such as the colouring of neighboring regions or the street network problem („travelling salesman" and its variants). Recall that in combinatorial topology, characteristic numbers (such as the winding number) of forms can be utilized to infer properties of the activity of suitable vector field actions defined on the respective forms.

Two more polynomials are of importance for our purpose here: One is the *Jones polynomial*, called $V_K(t)$, essentially a Laurent series in $\sqrt{t}$ associated to some oriented knot K such that

(1) if K is neighborhood isotopic to K', then $V_K(t) = V_{K'}(t)$,

(2) $V(\otimes) = 1$ (self-loop),

(3) $t^{-1}V_X - t\,V_X = (\sqrt{t} - (1/\sqrt{t}))\,V$ (skein relations),

and such that in particular, $V_K(t) = L_K(t^{-1/4})$. Finally, we can also define the *Kauffman polynomial* $Y_K$ which is given such that $Y_K = a^{-\omega(K)} D_K$, and especially bears the relationship

$$V_K(t) = [Y_K(t^{-1/4} - t^{1/4}, -t^{-3/4})]/(-t^{-1/2} - t^{1/2}).$$

So what we have essentially done is to construct *knot invariants* by combinatorial means in analogy to *partition functions*. This can be read as a strict analogy, because the relevant expression of the form

$$\langle K \rangle = \sum_\sigma \langle K \mid \sigma \rangle \, \lambda^{\mid \sigma \mid}$$

describes an expectation value of the knot diagram, and the product on the right-hand side expresses a product of weights. The knot diagram can be visualized then as an observable for a system of states. In the continuous version in terms of path integrals, this can be generalized by defining the same on a set of gauge fields with values in a given Lie algebra, such that

$$Z_K = \int dA \, e^{\llcorner} \, \tau_K,$$

where $\tau_K = \mathrm{Tr} \, [P \exp \int_K A]$ is the trace of the path-ordered gauge exponential for the connection A. (This has been discussed first by Witten and Atiyah.) It can also be shown that spin networks can be retrieved by recalling that spinors of type $\psi^A$, A = 1,2 (complex), give rise to a group action of the form $U \in SL(2, C)$ which acts on a spinor such that

$$(U\psi)^A = \sum_B U^A{}_B \, \psi^A$$

with the conjugate spinor

$$\psi^*{}_A = \varepsilon_{AB} \, \psi^B; \quad \varepsilon_{AB} = \begin{pmatrix} 0 & 1 \\ -1 & 0 \end{pmatrix}$$

so that the scalar product is defined by

$$\psi\psi^* = \psi^A \, \varepsilon_{AB} \, \psi^B.$$

Hence, as to spin networks, we can visualize them alternatively, as simplices (and spin foams as complexes, respectively) which carry group representations on their edges and tensor products of them on their vertices.

We add the following two remarks: 1. The correspondance discussed before between spin networks and their dual 1-skeleta is reminiscent of the similar relationship between *Voronoi diagrams* and *Delaunay triangulations*. In fact, both of these pairs turn out to be equivalent. Indeed, by starting from a random distribution of points in a Voronoi fashion, it is likely that eventually classical space-time (or any other macroscopic geometry as to that) can be retrieved by means of „averaging" over spin networks, once a fine tuning can be achieved in the relationship between their respective dual 1-skeleta (the triangulations). Hence, the topological structure of space (visualized in terms of „combinatorial topology") can give hints as to the detailed structure of the spin foams actually under-lying and thereby producing it. Similarly, on a classical and local basis, namely within chemistry e.g., the analysis of the relationship of Voronoi diagrams and their associated Delaunay triangulations can be also utilized in order to infer from the topological structure of a molecular compound to the underlying enzyme activities actually producing it. Some results have been achieved already with respect to base pairing in DNA visualized in terms of Grover's quantum search algorithm by Patel. Earlier, Kauffman has discussed aspects of molecular folding utilizing knot theory. With the view to our attitude taken in the present paper, we would like to generalize this idea in discussing the correspondence between the macroscopic form of an urban structure as it can be described by an appropriate morphic language on the one hand, and the underlying microscopic processes actually producing it.

2. The other point concerns an idea of Lloyd's who advocates the concept of visualizing the Universe *altogether* as a quantum computer arguing that by its mere presence, the Universe is permanently storing and processing information. In particular, he would like to understand as a chief result of this computation the emergence of decohering histories which by themselves evolve sufficiently complex structures. This aspect is very much on the line of our own argument. Hence, we can already recognize that knot structures turn out to establish an elegant and straightforward method to evaluate complex interactions. On the other hand, more to the point, and coming from the side of knot invariants as discussed above in the sense of (Lou) Kauffman, we can visualize the evaluation of the Jones polynomial as a generalized quantum amplitude, as Kauffman has shown recently: The braiding part of the polynomial coming from the polynomial's skein relations as displayed above, can then be construed as a quantum computation. Hence, *knot invariants show up as quantum computation*. Kauffman utilizes the following concept of a quantum computation: It consists basically in the application of a unitary transformation U to an initial *qunit* (not a qubit with two entries, but now with n entries) $\psi$ with $|\psi|^2 = 1$ plus an observation of $U\psi$. This will return the ket $|a>$ say, with probability $|U\psi|^2$. If we start in $|a>$, then the probability that this arrangement will return $|b>$ is in fact

$$|<b|U|a>|^2.$$

Then we introduce an operator formalism in Dirac notation:

$\cup$ (cup) = $|a>$ : $C \rightarrow V \otimes V$ (creation ket),

$\cap$ (cap) = $<b|$ : $V \otimes V \rightarrow C$ (annihilation bra),

$<b|o|a> = <b|a>$ : $C \rightarrow C$ (vacuum-vacuum amplitude).

Note that with Pr = │y><x│ as projection operator, and Q:= Pr/ <x│y>, we have also QQ = Q. That is, the sum of projections is equal to the identity which relates to the completeness of intermediate states. In other words: The amplitude for going from *a* to *b* consists of the summations of contributions from all the paths connecting *a* with *b*. This is obviously consistent with the Feynman picture discussed earlier. Hence, what we have to do is to visualize the bracket model as a vacuum-vacuum amplitude. Then it can be configured as a composition of operators (cups, caps, braiding) – provided the braiding is unitary. This can be clearly viewed as a quantum computation: Choose the Cup as „preparation" part and the Cap as „detection" part of the computing, and call M the unitary braiding operator, then this can be summarized by the expression

$$Z_K = <Cup│M│Cap>.$$

Hence, referring back from knots to their constituents, it is spin networks that can be thought of as representing the most *fundamental channels of information transport*: It is in fact *quantum computation* which is permanently being performed through the channels the spin networks provide. The latter serve as a kind of universal lattice through which the information produced by quantum computation is percolating such that an eventual threshold clustering in the sense of percolation theory spontaneously creates the onset of classicity. The actual route taken is that via the formation of knots in terms of spin networks and loops. That is, in the end, the classical world can be visualized equivalently as a „condensation" of knotted spin networks. More recently, a similar path (but with different conclusions) has been taken by Stuart Kauffman and Lee Smolin: They basically utilize a deterministic model of directed percolation to achieve similar results and show that this can be visualized as a cellular automaton. The idea is then to give the necessary criteria for a percolation phase transition which renders the system behaviour critical, which turns out to be essentially a derivation from Kauffman's idea of a „fourth law" of thermodynamics: He basically asks whether we can find a sense in which a non-ergodic Universe expands its total dimensionality, or „total work space", in a sustainable way as fast as it can. He then refers to Smolin's interpretation of spin networks and their knotted structures at Planck scale level as „comprising space itself". He suggests that knotted structures are combinatorial objects rather like molecules and symbol strings in grammar models, and he expects that such systems become „collectively autocatalytic" - practically showing up as *knots acting on knots to create knots* in rich coupled cycles not unlike a metabolism: „The connecting concept will be that those pathways into the adjacent possible along which the adjacent possible grows fastest will simultaneously be the most complex and most readily lead to quantum decoherence, and classicity. If complexity 'breeds' classicity, then the Universe may follow a path that maximizes complexity." This is in fact his con-cept of a „fourth law“: that the adjacent possible will be attained which could account for the explicit „semantics“ and dynamics of evolution.

There are two aspects in favour of these ideas: 1. the already mentioned case that classical geometry can be approximated by an average over all possible Voronoi configurations over a (n as yet) Riemannian manifold, as shown in the works of Luca Bombelli, and 2. the also mentioned case of the Barrett-Crane model for spin foams. As John Baez has formulated once in an e-mail discussion: „Now some people call a state in $C^2$ a ‚spinor‘, but other people call it a ‚qubit‘. And what we are really doing, from the latter viewpoint, is writing down ‚quantum logic gates‘ which manipulate ‚qubits‘ in an SU(2)-invariant way – in fact, an SL(2,C)-invariant way! In short, we're seeing how to build little Lorentz-invariant quantum computers. From this crazy viewpoint, what the Barrett-Crane model does is to build a theory of quantum gravity out of these little Planck-scale quantum computers.“

## 4 Conclusion

It should have become transparent after all what we are aiming at: The idea is to visualize the evolution of a macroscopically observable urban structure as the outcome of a superposition of social actions modelled in terms of a state function formalism (or decoherence) of all constitutive processes taking place on the microlevel of the respective social collective. The methods applied shall enable the inference from the observed topological structure to the underlying dynamics producing it. In other words: The social logic of space shall be retrieved by reconstructing the fundamental motion on which it is actually being based. *Process governs structure* in this sense, not viceversa, as has been assumed for a long time. The relation between process and structure resembles therefore the celebrated relation between micromotives and macrobehaviour in the sense of Thomas Schelling. As to *decentralization* being visualized as an organizing principle for urban structures in particular, we notice that this specific evolutionary mode is implemented into the systematics of the approach from the outset, because it turns out as a straightforward consequence of the mediation structure between micro and macro, in the first place. With a view to the actual research undertaken with respect to the evolution and structure of the historical centre of Bologna, the idea is to compare the specific logic of space in that case with the former political concept of decentralization as it has been operated by the city administration for a long while in the past. Further work is forthcoming.

## 5 Acknowledgements

For illustrative discussions and insight I thank my cooperation partners from Bologna, Anna Soci, and Giorgio Colacchio. In particular, I thank Anna Soci for inviting me to her institute. I also thank Wolfgang Watzlaw of the „Kulturoe-kologisches Forum Berlin" for his interest in this project and for giving me the occasion of delivering a number of talks in his group. Likewise, I thank the „Zentrum fuer interdisziplinaere Forschung" in Bielefeld and the organizer of the preparatory meetings to the research project „A General Theory of Information Transfer and Combinatorics", Rudolf Ahlswede, for hospitality and the occasion to give a conceptual talk on a similar topic of more fundamental scope.